\pdfoutput=1
\documentclass[11pt]{article}

\usepackage[T1]{fontenc}
\usepackage{lmodern}
\usepackage[utf8]{inputenc}
\usepackage[letterpaper,margin=1in]{geometry}
\usepackage{cite}
\usepackage{amsmath,amssymb,amsfonts}
\usepackage{graphicx}
\usepackage{ragged2e}
\usepackage{textcomp}
\usepackage{booktabs}
\usepackage{url}

\title{Evaluating AST-Derived Knowledge Graph Retrieval for AIOps Incident
Diagnosis in Apache Kafka}

\author{%
Shashank Mishra\thanks{Corresponding author: Shashank Mishra
(e-mail: \texttt{24shashankm@gmail.com}). This work was conducted as
independent research.}\\
Independent Researcher\\
Seattle, WA, USA\\
\texttt{24shashankm@gmail.com}
\and
Shweta Mishra\\
Independent Researcher\\
San Jose, CA, USA\\
\texttt{1196shweta@gmail.com}}

\date{September 9, 2026}

\begin{document}
\maketitle

\begin{abstract}
\justifying
Incident diagnosis requires retrieving source-linked operational evidence from large codebases. We evaluate graph-augmented retrieval for Apache Kafka using an abstract-syntax-tree (AST)-derived ownership graph over 1,600 source-linked log records and an audited benchmark of 315 incident-style queries covering 103 unique targets. Vector-only retrieval is a strong baseline at Recall@20 of 0.956, and neither static nor confidence-guided staged AST expansion significantly improves Recall@20 over it. As an external control we evaluate HippoRAG, a published text-derived graph-retrieval system, against its own ColBERTv2 retrieval backbone. Against ColBERTv2-only (Recall@20 0.803), HippoRAG attained 0.308; a post-hoc document-ensemble sensitivity raised Recall@20 to 0.648 but did not eliminate the deficit. Relative to static expansion, staging reduces cross-encoder pair workload and measured reranking time by about 52\%. A manually evaluated 100-query downstream study finds no significant root-cause-correctness advantage for AST-derived structural evidence.
\end{abstract}


\vspace{1em}

\section{Introduction}
\label{sec:introduction}

Debugging incidents in a distributed system is difficult because evidence can span components, source files, methods, and emitted log statements. For Apache Kafka, an AIOps system must first retrieve the correct source-linked evidence before a generator can offer a grounded diagnosis. Dense retrieval is a strong way to connect an incident description to a semantically related log template, but its ranking does not explicitly traverse the source ownership relations that connect a log statement to its method and class.

This paper evaluates whether those local relations are useful. We construct an AST-derived graph linking Kafka classes, methods, and emitted log templates. Retrieval-time expansion is deliberately narrow: a dense-retrieved log can contribute sibling logs from its owning method and enclosing class before cross-encoder reranking. We call this operation \emph{local AST-derived method/class graph expansion}. The terminology is intentionally specific; the study does not evaluate knowledge-graph retrieval in general.

Graph augmentation is not a single technique, however, and different systems derive their graphs from different signals. Our graph encodes program-structural ownership recovered deterministically from source syntax. Published text-derived systems such as HippoRAG \cite{b9} instead induce a graph from the text of the indexed units themselves, using LLM-based open information extraction and ranking documents by Personalized PageRank over the induced structure. Evaluating both families on the same benchmark helps determine whether an observed absence of graph benefit is specific to local source ownership or persists under a very different graph-construction strategy. We therefore add a controlled external baseline rather than relying on published numbers obtained on a different corpus and task.

We also evaluate confidence-guided staging. Static expansion reranks the full expanded pool, whereas staging incrementally expands and reranks candidates until a confidence rule stops or a fixed maximum is reached. Relative to static AST-KG, staging may reduce cross-encoder pair workload and measured cross-encoder reranking time even if structural expansion does not improve retrieval quality. The evaluation therefore separates exact-target retrieval quality, reranking efficiency, and downstream diagnostic utility.

The study addresses four research questions without presupposing positive answers:
\begin{itemize}
\item \textbf{RQ1:} Does AST-derived local structural expansion improve exact-target incident-log retrieval?
\item \textbf{RQ2:} How does confidence-guided staging affect reranking cost and retrieval quality relative to static AST-derived expansion?
\item \textbf{RQ3:} Does AST-derived structural evidence improve downstream incident diagnosis?
\item \textbf{RQ4:} How does a published text-derived graph-retrieval method transfer to this exact source-location task, and what incremental effect does its graph pipeline have relative to its own retrieval backbone?
\end{itemize}

We compare six primary retrieval conditions: vector-only retrieval, vector retrieval with cross-encoder reranking, static AST-derived expansion, confidence-guided staged AST-derived expansion, ColBERTv2-only retrieval, and HippoRAG with document ensembling disabled. We additionally report one post-hoc HippoRAG sensitivity condition with document ensembling enabled. The audited benchmark contains 315 queries over 103 unique targets. Vector-only and staged AST-KG each attain 301/315 hits at \(K=20\), static AST-KG attains 298/315, and Recall differences are not statistically significant. Vector-only has the highest MRR at every tested depth. Relative to static AST-KG, staging nevertheless reduces cross-encoder pair workload and measured cross-encoder reranking time substantially. A separate 100-query downstream evaluation finds no significant root-cause-correctness advantage for either AST-KG condition.

For the external baseline, the interpretation rests on the within-backbone contrast rather than on absolute scores. ColBERTv2 is HippoRAG's late-interaction retrieval backbone, and it differs from the \texttt{text-embedding-3-large} model used by our dense conditions; comparing HippoRAG directly against our dense baseline would therefore confound the graph pipeline with the choice of retriever. Running both conditions on the identical frozen corpus, direct ColBERTv2 retrieval attains Recall@20 of 0.803, whereas HippoRAG under its documented custom-dataset configuration attains 0.308. The paired difference of \(-49.5\) percentage points (exact two-sided McNemar \(p=2.18\times10^{-41}\)) is the primary external-baseline result, and it measures the incremental effect of HippoRAG's extraction, graph-construction, entity-linking, and Personalized PageRank pipeline relative to direct retrieval under the same backbone.

Because that reduction was unusually large, we additionally ran one post-hoc sensitivity analysis enabling HippoRAG's document-ensemble option, holding every other parameter and every frozen input fixed. Recall@20 rises to 0.648, yet ensemble-enabled HippoRAG remains 15.6 percentage points below its matched ColBERTv2-only control. The magnitude of the HippoRAG deficit is therefore configuration-sensitive, whereas its direction is unchanged across the two evaluated graph-ranking conditions.

The contributions are:
\begin{itemize}
\item An AST-derived Kafka structural graph containing 1,096 classes, 7,527 methods, and 1,600 source-owned log templates.
\item An audited 315-query, 103-target benchmark with complete AST, Neo4j, and Qdrant target coverage, while preserving the original 321-query archive unchanged.
\item A controlled comparison of four internal retrieval methods that isolates dense retrieval, cross-encoder reranking, local method/class expansion, and confidence-guided staging.
\item A controlled transfer evaluation of HippoRAG with a matched ColBERTv2-only control, together with a post-hoc document-ensemble sensitivity analysis, separating the external graph pipeline from its retrieval backbone and characterizing the configuration dependence of the result.
\item Paired, target-balanced, target-cluster-bootstrap, and \(K\)-sensitivity analyses that account for repeated queries per target.
\item Evidence that, relative to static AST-KG, confidence-guided staging reduces cross-encoder pair workload and measured cross-encoder reranking time while achieving 301/315 Hits@20.
\item A 100-query, unique-target downstream study with blinded manual evaluation finds no statistically reliable diagnostic improvement from AST-derived structural evidence.
\end{itemize}

The remainder of this paper is organized as follows. Section~\ref{sec:related} reviews related work. Section~\ref{sec:framework} describes the retrieval methods, including the external text-derived graph baseline. Section~\ref{sec:setup} presents the experimental setup. Section~\ref{sec:results} reports retrieval, external-baseline, graph-diagnostic, and downstream results. Section~\ref{sec:discussion} discusses their implications, Section~\ref{sec:threats} summarizes threats to validity, and Section~\ref{sec:conclusion} concludes.

\section{Related Work}
\label{sec:related}

Retrieval-augmented generation (RAG) grounds generated answers in external evidence retrieved from a non-parametric index \cite{b1}. Subsequent RAG variants have explored self-reflection, correction, adaptivity, active retrieval, trust calibration, retrieval pretraining, and retrieval-augmented few-shot learning \cite{b14,b15,b16,b17,b18,b19,b20,b21,b22}. Dense retrieval methods such as DPR \cite{b2} made this pattern practical at scale, but standard dense retrieval usually treats each indexed unit as an independent text chunk. This is limiting for incident debugging, where a relevant log statement is often meaningful only in relation to its source-code context, such as the method, class, or nearby emitted logs.

Graph-augmented retrieval methods use graph structure or graph-derived summaries to expand retrieval beyond isolated chunks \cite{b3,b9,b10}. KG-RAG, GraphRAG surveys, structured-data prompting, and graph retrieval under incomplete knowledge provide descriptive context for this design space \cite{b24,b25,b26,b27,b28}. Our approach is itself a graph-augmented retrieval method: it derives a structural graph directly from Kafka AST ownership and expands dense-retrieved log templates through their enclosing methods and classes. Published graph-retrieval frameworks differ chiefly in how the graph is obtained. HippoRAG \cite{b9} induces a schema-free graph from passage text using LLM-based open information extraction, links query entities into that graph with a late-interaction retriever \cite{b29,b35}, and ranks documents by Personalized PageRank; it reports gains on multi-hop question-answering benchmarks. Microsoft GraphRAG and GRAG \cite{b3,b10} likewise build graphs or graph-derived summaries from text rather than from program structure.

Among these, we evaluate HippoRAG experimentally on our Kafka benchmark, together with a matched retriever-only control, because its graph is induced from the text of the same indexed units and therefore offers a direct contrast with our AST-derived construction. GraphRAG and GRAG remain literature context and are not implemented as experimental baselines here. Our experiments consequently support conclusions about the evaluated AST-derived graph-expansion strategy relative to dense and cross-encoder baselines, and about the transfer of one published text-derived graph-retrieval method to this task, but not claims of superiority over graph-RAG approaches in general.

This paper also relates to log analytics and AIOps incident diagnosis. Prior work on log parsing and log benchmarks shows that reliable log-template extraction is important for downstream log analysis \cite{b4,b5}. Larger recent evaluations and surveys further show that automated log analysis remains sensitive to parser choice, system domain, and benchmark design \cite{b31,b32}. AIOps benchmark and survey work further emphasizes the need for realistic incident-management tasks and reproducible evaluation settings \cite{b11,b12}. Industrial incident-analysis work has also shown the value of structured root-cause knowledge for operational debugging \cite{b6}. In contrast to anomaly-detection tasks, our benchmark evaluates whether a natural-language incident query can retrieve the correct source-linked log evidence.

Finally, our retrieval pipeline builds on code-aware retrieval and cross-encoder reranking. Code-oriented models such as CodeBERT show the value of aligning natural language with program artifacts \cite{b7}. More broadly, dense retrieval without explicit relevance labels, late-interaction retrieval, and learning-to-rank methods provide complementary ranking perspectives for retrieval systems \cite{b23,b29,b30,b33}. Cross-encoder rerankers provide a second-stage scoring mechanism by evaluating each query-candidate pair more directly than dense similarity alone \cite{b8,b13}. Here, the cross-encoder is used both after vector retrieval and after local AST-derived expansion, separating reranking from the addition of source-ownership neighbors. The cited literature motivates the comparison but does not imply superiority of the evaluated method.

\section{Method}
\label{sec:framework}

\begin{figure}[t]
\centering
\includegraphics[width=\linewidth]{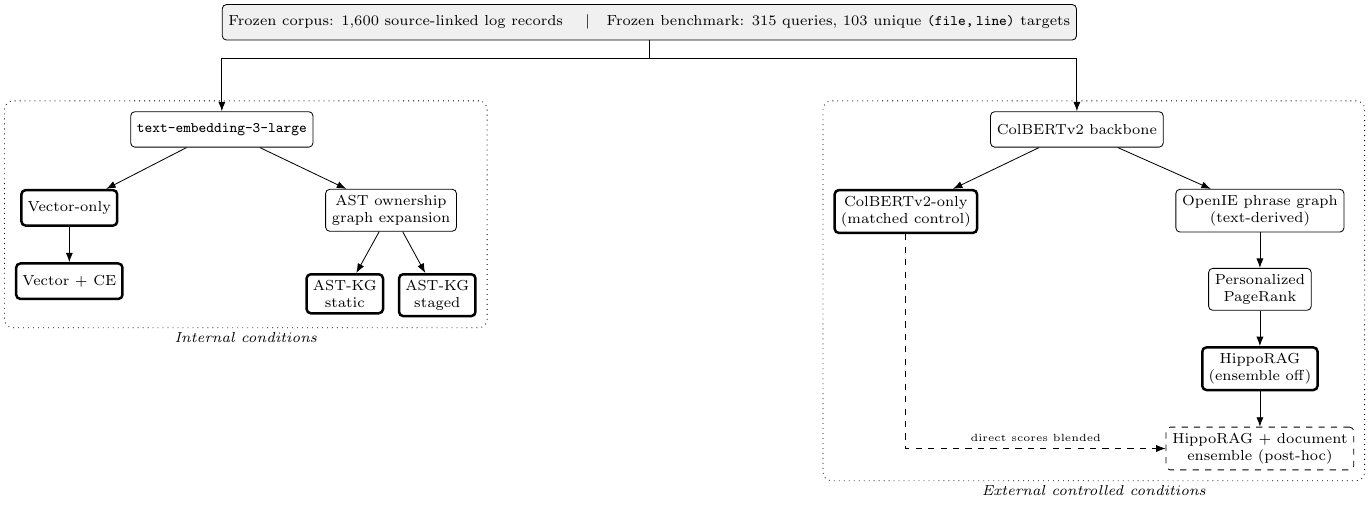}
\caption{Controlled experimental design. All conditions consume the identical frozen corpus and query set and are scored by the same exact \texttt{(file,\,line)} criterion. The internal conditions share the \texttt{text-embedding-3-large} dense representation and differ in reranking and AST-derived ownership expansion. The external conditions share the ColBERTv2 backbone, so the contrast between direct ColBERTv2 retrieval and HippoRAG estimates the incremental effect of the full text-derived graph pipeline rather than the choice of retriever. The two graph constructions differ in origin: AST ownership is recovered from program structure, whereas the HippoRAG graph is induced from record text by open information extraction. The dashed node marks the post-hoc document-ensemble sensitivity configuration.}
\label{fig:design}
\end{figure}

Figure~\ref{fig:design} summarizes the controlled design and the relationship between the internal and external conditions. The system constructs a lightweight AST-derived structural graph over Kafka source artifacts with the following node types:
\begin{itemize}
\item \texttt{Class}
\item \texttt{Method}
\item \texttt{LogTemplate}
\end{itemize}

Edges connect classes to declared methods and methods to emitted log templates. Retrieval uses only these ownership relations. A semantically retrieved log can expand to sibling logs emitted by the same method and then to logs owned by other methods of the enclosing class. Log templates are embedded in Qdrant for dense retrieval; Neo4j stores the ownership graph. Accordingly, the evaluated graph operation is local AST-derived method/class graph expansion, not a general test of KG-RAG.

\subsection{Source Parsing and Graph Construction}

We use a Tree-sitter AST parser for selected Java and Scala source areas instead of the earlier regular-expression implementation. The parser visits syntax-tree declarations and logger-invocation nodes, extracting package names, class-like declarations, methods, constructors, Scala \texttt{def} declarations, and supported log calls. Java classes, interfaces, enums, records, nested declarations, and anonymous classes are represented; Scala classes, objects, traits, enums, nested declarations, and functions are also represented. Method identifiers are normalized into signatures from AST parameter information. Supported logger tokens are \texttt{log}, \texttt{logger}, and \texttt{LOG}; supported levels are \texttt{trace}, \texttt{debug}, \texttt{info}, \texttt{warn}, \texttt{error}, and \texttt{fatal}. The first argument must be a string literal or a supported Scala string form.

AST ownership is explicit: each detected log template is attached to its enclosing syntax-tree method or function, and that owner is attached to its enclosing class-like declaration. Calls without an enclosing owner are skipped instead of being assigned to the nearest preceding method, as the legacy regex implementation could do heuristically. The graph schema is:
\begin{itemize}
\item \texttt{Class} properties: \texttt{fqn}, \texttt{name}, \texttt{file}, and \texttt{repo}.
\item \texttt{Method} properties: \texttt{fqn}, \texttt{class\_fqn}, \texttt{signature}, \texttt{name}, \texttt{file}, \texttt{line}, and \texttt{repo}.
\item \texttt{LogTemplate} properties: \texttt{repo}, \texttt{file}, \texttt{line}, \texttt{level}, \texttt{template}, and \texttt{logger}.
\item \texttt{(Class)-[:DECLARES]->(Method)}
\item \texttt{(Method)-[:EMITS]->(LogTemplate)}
\end{itemize}

The frozen corpus contains 1,096 \texttt{Class} nodes, 7,527 \texttt{Method} nodes, and 1,600 \texttt{LogTemplate} nodes, connected by 7,527 \texttt{DECLARES} edges and 1,600 \texttt{EMITS} edges. The reported experiments use the AST materialization stored in Neo4j and the Qdrant collection \texttt{kafka\_log\_templates\_ast}. These counts describe the configured extraction scope; because no independently labeled parser gold set was created, they do not establish comprehensive parser accuracy over all Kafka logging syntax.

\subsection{Dense Indexing and Retrieval Units}

Each \texttt{LogTemplate} node is converted into a dense-retrieval text record by concatenating the log level, normalized template string, logger name, method identifier, class identifier, and file:line location. Embeddings are generated with \texttt{text-embedding-3-large} and stored in Qdrant. This means that dense retrieval is performed over source-linked log templates rather than over raw files or arbitrary text chunks. The expected retrieval target for every benchmark question is therefore a specific \texttt{(file, line)} pair associated with one indexed log-template node.

We evaluate four internal retrieval modes:
\begin{enumerate}
\item \textbf{Vector-only:} semantic top-\(K\) retrieval over embedded log templates.
\item \textbf{Vector + Cross-Encoder:} vector retrieval followed by reranking.
\item \textbf{AST-KG static:} vector retrieval, local AST-derived method/class graph expansion, reranking of the expanded pool, and retention of the top \(K\).
\item \textbf{AST-KG staged:} incremental local expansion and reranking until the score-based confidence rule stops or the configured maximum pool size is reached.
\end{enumerate}
Two further external conditions, described in Section~\ref{subsec:external-method}, evaluate a published text-derived graph-retrieval method over the same retrieval units.

The reranking model is \texttt{BAAI/bge-reranker-v2-m3}. For static AST-KG, we first retrieve the top-\(K\) dense candidates, expand each seed through method and class ownership neighbors, deduplicate by \texttt{(file, line)}, rerank the expanded pool, and retain the top \(K\) final records. The staged method applies the same local expansion to an incrementally growing subset of dense seeds.

For the confidence-guided local AST-derived expansion policy, candidates in the current pool are scored by the cross-encoder and sorted using raw reranker scores. After each expansion step, the system computes an aggregate confidence value from four score-distribution signals: the relative margin between the top two scores, the relative drop from the top score to the lowest score in the current pool, the entropy of a softmax distribution over the scores, and a normalized variance term over the score distribution. All four confidence components are normalized to the unit interval before averaging; entropy is normalized by the maximum entropy of the current candidate pool, and the variance term is scaled relative to the top reranker score. The final stage confidence is the average of these four terms. Expansion stops when this aggregate confidence meets or exceeds \texttt{confidence\_high}, when no additional candidates are available, or when the maximum pool size \texttt{dynamic\_k\_max} has been reached. If confidence is at or below \texttt{confidence\_low}, the candidate pool is expanded more aggressively; otherwise, it is expanded with a smaller step. This policy allows the system to spend additional graph expansion and reranking work only when the current candidate pool is not yet sufficiently separated.

The downstream study uses the first, third, and fourth modes. The Vector+Cross-Encoder-only condition is absent from that study and is treated as a limitation rather than inferred from the retrieval results.

\subsection{External Text-Derived Graph Baseline}
\label{subsec:external-method}

To test whether the absence of a graph benefit is specific to AST-derived construction, we evaluate HippoRAG \cite{b9} on the identical retrieval universe, together with a matched retriever-only control. Both external conditions operate over exactly the same 1,600 source-linked records, serialized with exactly the semantic fields used by the dense baseline, and answer exactly the same 315 queries under the same exact \texttt{(file, line)} gold criterion.

\textbf{ColBERTv2-only.} A query is scored directly against all 1,600 records by ColBERTv2 late interaction \cite{b29,b35}, and documents are ranked by that score. No open information extraction, no graph construction, no query entity recognition, no Personalized PageRank, no graph expansion, and no cross-encoder reranking are applied. ColBERTv2 is HippoRAG's late-interaction retrieval backbone; it is not the embedding model used by our dense conditions, which use \texttt{text-embedding-3-large}.

\textbf{HippoRAG (ColBERTv2).} At indexing time, each of the same 1,600 records is passed through LLM-based named-entity recognition and open information extraction, producing entity and triple sets from which a phrase graph is assembled; synonymy edges are added between phrases whose ColBERTv2 similarity exceeds a threshold. At query time, entities are extracted from the query, linked to graph nodes using ColBERTv2, and Personalized PageRank is run from the linked nodes; the resulting node probabilities are propagated through fact and document incidence matrices to produce the document ranking.

Because both conditions share the same retrieval backbone, the same checkpoint, and the same document text, the contrast between them is not confounded by the choice of retriever. The primary controlled comparison therefore evaluates the incremental effect of HippoRAG's extraction, graph-construction, entity-linking, and Personalized PageRank pipeline taken together, relative to direct document retrieval under that backbone. It should not be read as isolating the contribution of graph edges alone, since several operations change simultaneously between the two conditions.

\textbf{Document ensembling.} HippoRAG exposes a \texttt{doc\_ensemble} option that changes how the final document score is formed. With the option disabled, which is the setting its custom-dataset documentation prescribes and the setting used for our primary condition, the ranking is the Personalized PageRank document distribution alone. With the option enabled, the implementation selects among three behaviours per query: if no entities are extracted from the query, documents are ranked by the direct ColBERTv2 score; if every linked entity scores above \texttt{recognition\_threshold}, the PageRank distribution alone is used; otherwise the two are blended in equal parts after min-max normalization of the direct scores. Enabling the option therefore does not simply add a term, but conditionally routes each query between graph-only, retriever-only, and blended ranking.

After observing the unusually large degradation under the primary configuration, we ran one post-hoc sensitivity analysis enabling this option while holding all other parameters and all frozen artifacts fixed. That analysis is reported separately in Section~\ref{subsec:external-results} and is not part of the pre-specified analysis plan.

\section{Experimental Setup}
\label{sec:setup}

\subsection{Benchmark Construction}

The final retrieval benchmark contains 315 Kafka incident-style queries assembled from three fixed repository artifacts. The \texttt{v2} subset contributes 150 queries generated from 50 sampled log-template targets. The AST-cleaned \texttt{v3} subset contributes 144 queries, with sampling focused on \texttt{WARN}/\texttt{ERROR} templates and balanced across selected Kafka source areas. The JIRA-seeded subset contributes 21 issue-linked queries. The 21 JIRA-seeded queries cover five unique expected source targets, so this subset should not be interpreted as 21 independent source-location cases. The benchmark therefore contains repeated descriptions of some targets and covers 103 unique expected \texttt{(file, line)} pairs.

The benchmark is designed to evaluate source-linked incident-log retrieval against known expected file/line targets. The \texttt{v2} and \texttt{v3} questions are generated from indexed \texttt{LogTemplate} records, while the JIRA-seeded subset adds issue-backed prompts with public issue metadata. Each query is paired with the source file and line number of the log template that motivated the question. Because multiple incident-style questions can target the same log template, the benchmark evaluates whether each retrieval method can identify the same source-linked evidence from different natural-language descriptions.

For reproducibility, the original merged benchmark is archived unchanged with 321 queries and 105 unique targets. The AST audit identified six queries whose expected targets were legacy-regex false positives rather than logger calls: \texttt{L081-Q1}, \texttt{L081-Q2}, and \texttt{L081-Q3} at \path{ResultOrError.java:89}, and \texttt{L084-Q1}, \texttt{L084-Q2}, and \texttt{L084-Q3} at \path{ResultOrError.java:87}. Both targets occur under \path{metadata/src/main/java/org/apache/kafka/controller/}. Removing only those six queries yields the final 315-query, 103-target benchmark. The audit reports zero missing expected targets in the source AST extraction, Neo4j graph, and Qdrant index. Deterministic identifiers and the preserved original archive make this correction traceable rather than a silent relabeling.

\subsection{Sampling for the Downstream Study}

For the downstream explanation study, we sample a 100-query subset from the merged 315-query benchmark using a fixed random seed (\texttt{sample\_seed = 42}). To avoid overrepresenting repeated prompts for the same log template, we set \texttt{allow\_same\_target = false} and keep at most one query per unique expected \texttt{(file, line)} target. Operationally, we generate a seeded random ordering of the merged benchmark and keep the first query for each previously unseen expected \texttt{(file, line)} target until 100 unique targets are collected.

For each sampled query, each evaluated method returns final-ranked evidence records. The fixed generator snapshot is \path{gpt-4.1-mini-2025-04-14} \cite{b34}; it receives the top five records. Each generated answer is manually scored for root-cause correctness, evidence grounding, and hallucination using a fixed binary rubric. Method identities are hidden behind neutral Method A/B/C labels during scoring. This study is a secondary diagnostic evaluation of whether retrieved evidence translates into more faithful explanations.

\subsection{Retrieval Configuration}

The paper configuration uses a retrieval pool size of \(K = 20\), reranker \texttt{BAAI/bge-reranker-v2-m3}, staged-confidence parameters \texttt{dynamic\_k\_min = 10}, \texttt{dynamic\_k\_max = 40}, \texttt{dynamic\_k\_drop = 0.15}, and \texttt{staged\_batch\_size = 20}, plus confidence thresholds of 0.60 and 0.25. The retrieval results for this main configuration are reported in Table~\ref{tab:retrieval-results}.

In addition to the main \(K=20\) benchmark, we ran a limited retrieval-depth sensitivity study over \(K \in \{5,10,15,20\}\) while keeping the other staged-confidence settings fixed at \texttt{dynamic\_k\_min = 5}, \texttt{dynamic\_k\_drop = 0.15}, \texttt{dynamic\_k\_max = 40}, \texttt{staged\_batch\_size = 20}, \texttt{confidence\_high = 0.60}, and \texttt{confidence\_low = 0.25}. The corresponding ablation results are reported in Table~\ref{tab:k-sensitivity}.

The equal weighting of the four confidence components and the 0.60/0.25 thresholds are heuristic design choices; this study does not establish their optimality, and systematic threshold sensitivity is left for future work.

\subsection{External Baseline Configuration}
\label{subsec:external-setup}

The six primary conditions comprise four internal conditions, namely vector-only, vector plus cross-encoder, static AST-KG, and staged AST-KG, and two external controlled conditions, namely ColBERTv2-only and HippoRAG with \texttt{doc\_ensemble = false}. A seventh condition, HippoRAG with \texttt{doc\_ensemble = true}, is a post-hoc sensitivity analysis rather than a primary condition, and is labelled as such wherever it appears.

The external conditions require the retriever-only control for a specific reason. HippoRAG uses ColBERTv2 as its late-interaction retrieval backbone, whereas our dense conditions embed records with \texttt{text-embedding-3-large}. A direct comparison between HippoRAG and vector-only retrieval would therefore confound the effect of the graph pipeline with the effect of substituting one retrieval system for another. Evaluating ColBERTv2-only on the identical corpus controls for the external method's backbone, so that the difference between ColBERTv2-only and HippoRAG with document ensembling disabled estimates the incremental effect of HippoRAG's full text-derived graph retrieval pipeline, comprising open information extraction and graph construction, entity recognition and linking, and Personalized PageRank document ranking, taken together. It does not isolate the contribution of graph edges alone.

We evaluate HippoRAG v1.0.0 because it corresponds to the published ColBERTv2-based HippoRAG method used in our matched-backbone comparison. HippoRAG 2 introduces materially different passage-integration and reasoning mechanisms and is therefore treated as a separate system outside the scope of this controlled evaluation. We use the official implementation released for the original HippoRAG work \cite{b9} at tag \texttt{v1.0.0}, commit \texttt{b144c46df14cabe5f5822d8caded4bec5f709461}, with the configuration its documentation prescribes for a custom dataset: ColBERTv2 retrieval backbone, \texttt{gpt-3.5-turbo-1106} for open information extraction and query entity recognition, synonym similarity threshold 0.8, Personalized PageRank damping 0.5, \texttt{graph\_alg = ppr}, \texttt{doc\_ensemble = false}, \texttt{node\_specificity = true}, \texttt{recognition\_threshold = 0.9}, and \texttt{max\_steps = 1}. The retriever-only control is the same implementation with \texttt{dpr\_only = true}, which bypasses query entity recognition, graph loading, and Personalized PageRank. No HippoRAG or ColBERTv2 hyperparameter was tuned on the benchmark, and the corpus and queries are byte-identical to those used by the internal conditions.

Before any HippoRAG retrieval result was observed, we fixed a pre-result analysis plan in the repository specifying the extraction and graph diagnostics to be reported, the metric definitions and statistical tests, and the interpretation branches to be used. This is an internal pre-specified diagnostic protocol recorded alongside the code, not a timestamped public preregistration.

The document-ensemble condition was run afterwards, in response to the observed result, and is therefore a post-hoc sensitivity analysis. Its only intended configuration change is \texttt{doc\_ensemble} from \texttt{false} to \texttt{true}. All frozen inputs are reused unchanged: the open information extraction output, the phrase graph and its synonymy edges, the cached query entity recognition, the 1,600-document corpus serialization, the 315 query strings, and the gold \texttt{(file, line)} labels. No extraction was re-run, no API call was made, and no threshold was tuned.

The experiment used the official algorithm and configuration with platform and data-loading adaptations required to run it in our environment. Execution was on Apple silicon under macOS on CPU, using \texttt{faiss-cpu} in place of \texttt{faiss-gpu} and Python 3.10.21 rather than 3.9; extraction ran sequentially because the upstream multiprocessing path relies on Linux \texttt{fork} semantics to share an initialized client; LangChain dependencies, which upstream leaves unpinned, were advanced to the minimum mutually compatible versions that allow the released code to import; and a compatibility adapter preserved the frozen Kafka document serialization for the ColBERTv2-only and primary ensemble-off paths. The post-hoc ensemble-on path uses the upstream stock constructor; the resulting serialization-path sensitivity is quantified in Section~\ref{sec:threats}. These adaptations were not intended to alter retrieval or extraction semantics; minor platform-dependent numerical differences remain possible. For the dependency update in particular, the extraction protocol was not intentionally changed: the model, prompts, temperature, token limits, and requested output format were unchanged.

\subsection{Downstream LLM Evaluation}

The downstream generator is \path{gpt-4.1-mini-2025-04-14}. For each retrieval method, it receives the top five final-ranked evidence records. The prompt instructs the generator to use only the retrieved evidence and to return a JSON object containing a probable root cause, impacted component, likely fix direction, evidence ranks, and a concise explanation.

Each generated answer is then scored with a strict binary rubric:
\begin{itemize}
\item \textbf{RootCauseCorrectness = 1} only when the predicted cause clearly matches the question.
\item \textbf{EvidenceGrounding = 1} only when the explanation is supported by the retrieved logs.
\item \textbf{Hallucination = 1} when the answer introduces an unsupported causal claim.
\end{itemize}

The 300 generated answers were evaluated manually by a single evaluator using the fixed binary rubric above. Method identities were randomized independently per query behind neutral Method A/B/C labels and remained blinded during scoring; unblinding occurred only after all answers had complete scores. The evaluation contains one judgment per answer and does not include an independent second evaluator or inter-annotator agreement. The 100-query study should therefore be read as a focused, single-evaluator blinded manual evaluation under a fixed rubric.

\subsection{Retrieval Metrics}
\label{subsec:metrics}

Each query has one gold \texttt{(file, line)} target. Let \(r_i\) be its rank in the method's final-ranked output. We define the per-query indicator as
\begin{equation}
R_i(K)=\begin{cases}1,& r_i\leq K,\\0,& r_i>K\ \text{or the target is absent.}\end{cases}
\end{equation}
Thus Recall@\(K\) equals one for a query only when its final gold rank is at most \(K\), and aggregate Recall@\(K\) is the mean of \(R_i(K)\). We report Hits@\(K\) as the numerator of that mean. MRR@\(K\) is the mean of \(1/r_i\) for targets with \(r_i\leq K\), with zero assigned otherwise. We use MRR@20 for the primary evaluation and MRR@\(K\) for the sensitivity analysis.

Because the Vector + CE condition reranks a fixed top-\(K\) dense candidate set without adding or removing candidates, Recall@\(K\) is identical to Vector-only by construction. This condition therefore isolates ranking effects, and MRR rather than Recall@\(K\) is the informative metric for evaluating its reranking behavior. We accordingly report no Recall test between Vector-only and Vector + CE.

Vector-only retrieves 301/315 queries at \(K=20\), leaving 14 misses and therefore a maximum possible absolute Recall@20 improvement of \(14/315 = 4.44\) percentage points on this benchmark. This is a headroom observation about the metric ceiling, not a statistical power or minimum-detectable-effect claim; no power analysis was conducted.

\subsection{Statistical Methods}

Binary paired Recall and downstream metric differences are tested with exact two-sided McNemar tests, implemented as exact binomial tests over discordant query pairs. Marginal binary rates use 95\% Wilson score intervals. Paired differences in MRR and downstream aggregate rates are additionally summarized using percentile paired-bootstrap confidence intervals. Because the retrieval benchmark repeats queries for some source targets, retrieval uncertainty is also estimated with a target-cluster bootstrap that resamples the 103 unique expected \texttt{(file, line)} targets and retains the queries within each sampled cluster. Target-balanced metrics are computed by averaging outcomes within each target and then averaging the 103 targets equally. Bootstrap procedures use a fixed seed of 42 and 10,000 replicates where applicable. For retrieval, exact McNemar tests are reported as query-level paired analyses, while target-cluster bootstrap intervals explicitly account for repeated queries associated with the same source target; retrieval conclusions are therefore interpreted jointly with the target-cluster intervals. For the downstream study, which contains 100 unique targets, exact McNemar tests serve as the primary paired binary analysis. Bootstrap intervals need not invert the discrete exact test.

\section{Results}
\label{sec:results}

\subsection{Retrieval Benchmark}

Table~\ref{tab:retrieval-results} summarizes the primary retrieval benchmark over all 315 incident-style queries.

\begin{table}[htbp]
\caption{Main \(K=20\) retrieval results on 315 queries and 103 unique targets. Six conditions are primary: four internal ones (Vector-only, Vector + CE, AST-KG static, AST-KG staged) and two external ones sharing the ColBERTv2 retrieval backbone (ColBERTv2-only and HippoRAG with document ensembling disabled). The remaining row is a single post-hoc sensitivity condition and is not a primary condition. Savings apply only to staged AST-KG and are relative to static AST-KG. Bold marks the best exact-target MRR@20 and the cross-encoder savings; Hits@20 and Recall@20 are left unemphasised because three conditions tie at 301/315 and Vector + CE ties with Vector-only by construction.}
\label{tab:retrieval-results}
\centering
\resizebox{\ifdim\width>\linewidth\linewidth\else\width\fi}{!}{%
\begin{tabular}{lrrrrr}
\toprule
Method & Hits@20 & Recall@20 & MRR@20 & CE pair save vs. static & CE-time save vs. static \\
\midrule
Vector-only & 301/315 & 0.956 & \textbf{0.751} & --- & --- \\
Vector + CE & 301/315 & 0.956\(^{\ddagger}\) & 0.697 & --- & --- \\
ColBERTv2-only & 253/315 & 0.803 & 0.552 & --- & --- \\
HippoRAG (ensemble off) & 97/315 & 0.308 & 0.110 & --- & --- \\
HippoRAG + document ensemble\(^{\dagger}\) & 204/315 & 0.648 & 0.219 & --- & --- \\
AST-KG static & 298/315 & 0.946 & 0.705 & --- & --- \\
AST-KG staged & 301/315 & 0.956 & 0.697 & \textbf{51.8\%} & \textbf{51.4\%} \\
\bottomrule
\end{tabular}%
}

\vspace{2pt}
{\footnotesize \(\dagger\) Post-hoc sensitivity configuration; the pre-specified HippoRAG condition is the ensemble-off row above. See Section~\ref{subsec:external-results}.\\
\(\ddagger\) Recall@\(K\) is identical to Vector-only by construction; see Section~\ref{subsec:metrics}.\par}
\end{table}

Static AST-KG is 0.95 percentage points below vector-only Recall@20, and staged AST-KG is equal to vector-only. Neither difference is significant. Target-balanced Recall@20 is 0.955 for vector-only, Vector + CE, and staged AST-KG, and 0.954 for static AST-KG. Target-balanced MRR@20 is 0.758, 0.703, 0.711, and 0.703, respectively, preserving the query-weighted interpretation. For the external conditions, target-balanced Recall@20 is 0.809 for ColBERTv2-only, 0.314 for HippoRAG with ensembling off, and 0.658 for the post-hoc ensembling-on condition, with target-balanced MRR@20 of 0.556, 0.112, and 0.223 respectively; target balancing therefore does not change the external ordering either.

\subsection{External Text-Derived Graph Baseline}
\label{subsec:external-results}

All external conditions share the ColBERTv2 backbone, the frozen 1,600-record corpus, the 315 queries, and the exact \texttt{(file, line)} gold criterion. Differences among them therefore estimate the incremental effect of HippoRAG's full text-derived graph retrieval pipeline, namely open information extraction and graph construction, entity recognition and linking, and Personalized PageRank document ranking, rather than the contribution of graph edges alone. We report the primary pre-specified comparison first, then the post-hoc sensitivity analysis.

Direct ColBERTv2 retrieval obtains 253/315 Hits@20, corresponding to Recall@20 of 0.803 and MRR@20 of 0.552.

Our initial HippoRAG evaluation used the documented custom-dataset configuration with document ensembling disabled and obtained Recall@20 of 0.308 (97/315 Hits@20, MRR@20 of 0.110), a reduction of \(-49.5\) pp relative to the matched backbone (95\% target-cluster interval \([-56.8, -42.1]\) pp; exact two-sided McNemar \(p=2.18\times10^{-41}\); discordance 5 HippoRAG-only against 161 ColBERTv2-only). Because the degradation was unusually large, we conducted a single post-hoc sensitivity analysis enabling document ensembling while holding the frozen corpus, graph, queries, and all other parameters fixed. Recall@20 rose to 0.648 (204/315 Hits@20, MRR@20 of 0.219), but remained below the matched ColBERTv2-only control at 0.803.

The gain over the ensemble-off condition is \(+34.0\) pp (95\% target-cluster interval \([+27.7, +40.4]\) pp; exact two-sided McNemar \(p=1.23\times10^{-32}\); discordance 107 ensemble-on-only against 0), with a paired MRR@20 gain of \(+0.109\) (\([+0.089, +0.131]\)) and no losses. Under this condition the implementation routed 119 queries through the PageRank-only path, 192 through the blended path, and 4 through the direct-retrieval path after no query entities were extracted. Relative to the backbone, ensemble-enabled HippoRAG was 15.6 percentage points below its matched ColBERTv2-only control (95\% target-cluster interval \([-21.2, -10.2]\) pp; exact two-sided McNemar \(p=1.36\times10^{-10}\); discordance 7 ensemble-on-only against 56 ColBERTv2-only), with a paired MRR@20 difference of \(-0.333\) (\([-0.396, -0.268]\)).

Document ensembling therefore substantially mitigated the HippoRAG transfer degradation, but did not eliminate it. Under the ensemble-enabled sensitivity configuration, Recall@20 increased from 0.308 to 0.648, yet ensemble-enabled HippoRAG remained 15.6 percentage points below its matched ColBERTv2-only control. Thus, the magnitude of the observed deficit is configuration-sensitive, while its direction remained unchanged under the evaluated sensitivity condition. This scoping applies to the frozen Kafka benchmark, its short source-linked operational records, the exact \texttt{(file, line)} retrieval objective, and the HippoRAG v1 configuration family evaluated here. The ordering is stable across depth, with ColBERTv2-only \(>\) HippoRAG ensemble-on \(>\) HippoRAG ensemble-off at every tested \(K\), as reported in Table~\ref{tab:k-sensitivity}.

A separate and distinct comparison concerns the two retrieval systems themselves: ColBERTv2-only was 15.2 percentage points below Vector-only at Recall@20 (\([-20.8, -10.2]\) pp, \(p=9.06\times10^{-14}\)). This backbone gap should not be confused with the 15.6-point gap between ensemble-enabled HippoRAG and ColBERTv2-only reported above; the two happen to be similar in size but compare different pairs of systems. This backbone gap motivates the retriever-only control: because the two retrieval systems differ substantially, the within-backbone contrast between HippoRAG and ColBERTv2-only is the appropriate basis for interpreting the incremental effect of the external graph pipeline as a whole.

\subsection{External Baseline Graph Diagnostics}
\label{subsec:diagnostics}

The diagnostics in Table~\ref{tab:graph-diagnostics} were specified before any HippoRAG retrieval result was observed, and are reported before interpreting retrieval quality.

\begin{table}[htbp]
\caption{Extraction and graph diagnostics for the HippoRAG index over the frozen 1,600-document Kafka corpus.}
\label{tab:graph-diagnostics}
\centering
\resizebox{\ifdim\width>\linewidth\linewidth\else\width\fi}{!}{%
\begin{tabular}{lr}
\toprule
Diagnostic & Value \\
\midrule
Documents & 1,600 \\
Entity occurrences extracted & 10,522 \\
Unique entity strings & 2,787 \\
Entities per document (mean / median) & 6.58 / 6 \\
Triples extracted & 9,941 \\
Unique facts & 8,785 \\
Triples per document (mean / median) & 6.21 / 6 \\
Documents yielding no entities & 82 (5.1\%) \\
Documents yielding no relations & 95 (5.9\%) \\
\midrule
Graph phrase nodes & 3,603 \\
Graph edges & 118,870 \\
Graph density & 0.0183 \\
Mean / median node degree & 66 / 15 \\
Connected components & 2 \\
Largest component & 3,602 nodes (99.97\%) \\
Isolated (degree-zero) nodes & 1 \\
\bottomrule
\end{tabular}%
}
\end{table}

The extracted graph was neither sparse nor fragmented: 94.9\% of documents yielded at least one entity, 94.1\% yielded at least one relation, and 99.97\% of phrase nodes belonged to the largest connected component, leaving a single isolated node. The result is therefore not readily explained by widespread extraction sparsity or graph fragmentation. However, connectivity does not establish semantic specificity or usefulness, and the graph is not dense in the graph-theoretic sense: its edge density is 0.0183.

Two further observations bound alternative explanations. Only 4 of 315 queries yielded no entities and thus fell back to an uninformative uniform document ranking, so that fallback does not account for the aggregate; restricting attention to the 311 entity-bearing queries leaves 97 hits. Hits@20 is also approximately flat in the number of extracted query entities, at 0.339, 0.311, 0.276, 0.317, and 0.250 for queries with one, two, three, four, and five or more entities, so query-side entity sparsity does not account for the result either.

Ranking concentration over the 1,600-document corpus differs across the external conditions. Counting distinct documents appearing anywhere in the 315 top-20 lists, ColBERTv2-only covers 1,080 documents, HippoRAG with ensembling disabled covers 844, and the ensemble-enabled condition covers 1,036. The most frequently returned document appears in 54, 123, and 87 of the 315 lists respectively, and the most frequent top-one document repeats 9, 23, and 12 times. Document ensembling was therefore associated with greater document coverage and lower ranking concentration, moving these statistics toward the direct ColBERTv2 retrieval condition.

The extracted vocabulary offers a descriptive counterpart. Auditing the five most frequent extracted phrases against the frozen serialization, \texttt{logger}, \texttt{class} and \texttt{location} are literal field keys present in all 1,600 records; \texttt{debug} is the value of the level field in 567 records; and \texttt{error} is both the level value in 272 records and generic logging vocabulary occurring in the message text of a further 249. The most frequent extracted phrases were thus dominated by record-schema metadata and generic logging vocabulary rather than Kafka-specific entities, although domain phrases such as \texttt{topicpartition} (197 occurrences), \texttt{partition} (140) and \texttt{node} (123) also occur frequently among the 2,787 unique extracted strings. This suggests that open information extraction over short key--value operational records partially captures serialization structure and logging terminology in addition to domain content. The pattern is consistent with graph propagation through high-frequency generic phrases influencing ranking concentration; our experiments do not isolate this mechanism as the sole cause.

\subsection{K-Sensitivity Ablation}

Table~\ref{tab:k-sensitivity} reports a limited retrieval-depth ablation over \(K \in \{5,10,15,20\}\) while holding the other staged-confidence settings fixed.

\begin{table}[htbp]
\caption{Retrieval-depth sensitivity on 315 queries, internal and external conditions. Savings are relative to static AST-KG at the same \(K\); dashes denote not applicable, and no external condition uses a cross-encoder. The internal sensitivity run used \texttt{dynamic\_k\_min=5}, whereas the primary \(K=20\) configuration used \texttt{dynamic\_k\_min=10}; the two runs report 53.4\% and 51.4\% measured cross-encoder-time saving at \(K=20\) respectively. \(\dagger\) Post-hoc sensitivity configuration.}
\label{tab:k-sensitivity}
\centering
\scriptsize
\resizebox{\ifdim\width>\linewidth\linewidth\else\width\fi}{!}{%
\begin{tabular}{rlrrrr}
\toprule
\(K\) & Method & Recall@\(K\) & MRR@\(K\) & CE pair save vs. static & CE-time save vs. static \\
\midrule
5 & Vector-only & 0.860 & \textbf{0.742} & --- & --- \\
5 & Vector + CE & 0.860 & 0.708 & --- & --- \\
5 & AST-KG static & 0.835 & 0.687 & --- & --- \\
5 & AST-KG staged & 0.838 & 0.694 & 59.0\% & 58.3\% \\
5 & ColBERTv2-only & 0.667 & 0.538 & --- & --- \\
5 & HippoRAG (ens. off) & 0.190 & 0.099 & --- & --- \\
5 & HippoRAG + ens.\(^{\dagger}\) & 0.413 & 0.194 & --- & --- \\
\midrule
10 & Vector-only & 0.908 & \textbf{0.748} & --- & --- \\
10 & Vector + CE & 0.908 & 0.705 & --- & --- \\
10 & AST-KG static & 0.902 & 0.697 & --- & --- \\
10 & AST-KG staged & 0.908 & 0.704 & 62.0\% & 62.7\% \\
10 & ColBERTv2-only & 0.749 & 0.548 & --- & --- \\
10 & HippoRAG (ens. off) & 0.244 & 0.106 & --- & --- \\
10 & HippoRAG + ens.\(^{\dagger}\) & 0.533 & 0.211 & --- & --- \\
\midrule
15 & Vector-only & 0.940 & \textbf{0.750} & --- & --- \\
15 & Vector + CE & 0.940 & 0.703 & --- & --- \\
15 & AST-KG static & 0.924 & 0.702 & --- & --- \\
15 & AST-KG staged & 0.933 & 0.701 & 58.7\% & 56.3\% \\
15 & ColBERTv2-only & 0.787 & 0.551 & --- & --- \\
15 & HippoRAG (ens. off) & 0.273 & 0.108 & --- & --- \\
15 & HippoRAG + ens.\(^{\dagger}\) & 0.597 & 0.216 & --- & --- \\
\midrule
20 & Vector-only & 0.956 & \textbf{0.751} & --- & --- \\
20 & Vector + CE & 0.956 & 0.697 & --- & --- \\
20 & AST-KG static & 0.946 & 0.705 & --- & --- \\
20 & AST-KG staged & 0.956 & 0.697 & 51.8\% & 53.4\% \\
20 & ColBERTv2-only & 0.803 & 0.552 & --- & --- \\
20 & HippoRAG (ens. off) & 0.308 & 0.110 & --- & --- \\
20 & HippoRAG + ens.\(^{\dagger}\) & 0.648 & 0.219 & --- & --- \\
\bottomrule
\end{tabular}%
}
\end{table}

No AST-KG Recall advantage is significant at \(K\in\{5,10,15,20\}\). Vector-only has the highest MRR@\(K\) at every depth, while staged AST-KG equals or exceeds static AST-KG Recall@\(K\) at every depth. Relative to static AST-KG at the same \(K\), staging also reduces cross-encoder pair workload and measured cross-encoder reranking time. The sensitivity experiment used \texttt{dynamic\_k\_min=5}; the primary configuration used \texttt{dynamic\_k\_min=10}. The \(K=20\) sensitivity run reports 53.4\% measured cross-encoder-reranking-time saving, compared with 51.4\% in the primary run. Both runs report a 51.8\% cross-encoder pair reduction at the reported precision. The primary-run artifacts retain the reported relative savings but not absolute cross-encoder pair counts or reranking seconds, whereas the sensitivity-run artifacts retain pair counts but not absolute reranking seconds. Consequently, the two runs' absolute reranking workloads and timings cannot be compared directly. Because wall-clock timing is also environment-dependent, we report both values without attributing the difference causally to the \texttt{dynamic\_k\_min} change. The stable qualitative findings show that the primary \(K=20\) conclusions are robust to retrieval depth. The external ordering is likewise stable: ColBERTv2-only exceeds both HippoRAG conditions at every tested depth in both Recall@\(K\) and MRR@\(K\), the ensemble-on condition exceeds the ensemble-off condition at every depth, and all three remain below the four internal conditions throughout.

\subsection{Retrieval Paired Analysis}

\begin{table}[htbp]
\caption{Paired retrieval statistics at \(K=20\). Comparisons are \(A-B\); discordant counts are \(A\)-only/\(B\)-only; Recall differences and their CIs are in percentage points. CIs are 95\% target-cluster bootstrap intervals over the 103 unique targets with 10,000 replicates and seed 42. McNemar \(p\)-values are query-level paired tests; confidence intervals account for repeated queries through target-cluster resampling. \(\dagger\) Post-hoc sensitivity configuration. \(\ddagger\) Vector + CE reranks a fixed dense candidate set, so its Recall@\(K\) is identical to Vector-only by construction and no Recall test is reported; MRR remains informative.}
\label{tab:retrieval-statistics}
\centering
\scriptsize
\resizebox{\ifdim\width>\linewidth\linewidth\else\width\fi}{!}{%
\begin{tabular}{llrrl}
\toprule
Comparison \(A-B\) & Recall diff. [95\% CI] (pp) & Discordant & Exact McNemar \(p\) & MRR@20 diff. [95\% CI] \\
\midrule
\multicolumn{5}{l}{\emph{Internal conditions}} \\
Vector + CE -- Vector-only & N/A\(^{\ddagger}\) & N/A\(^{\ddagger}\) & N/A\(^{\ddagger}\) & -0.054 [-0.099, -0.009] \\
AST-KG static -- Vector-only & -0.95 [-5.17, +2.54] & 4/7 & 0.549 & -0.047 [-0.087, -0.006] \\
AST-KG staged -- Vector-only & 0.00 [-0.95, +0.95] & 1/1 & 1.000 & -0.054 [-0.100, -0.008] \\
AST-KG staged -- AST-KG static & +0.95 [-2.56, +5.10] & 7/4 & 0.549 & -0.007 [-0.027, +0.011] \\
\midrule
\multicolumn{5}{l}{\emph{External conditions}} \\
HippoRAG + ens.\(^{\dagger}\) -- ColBERTv2-only & -15.6 [-21.2, -10.2] & 7/56 & \(1.36\times10^{-10}\) & -0.333 [-0.396, -0.268] \\
HippoRAG + ens.\(^{\dagger}\) -- HippoRAG (ens. off) & +34.0 [+27.7, +40.4] & 107/0 & \(1.23\times10^{-32}\) & +0.109 [+0.089, +0.131] \\
HippoRAG (ens. off) -- ColBERTv2-only & -49.5 [-56.8, -42.1] & 5/161 & \(2.18\times10^{-41}\) & -0.442 [-0.508, -0.375] \\
ColBERTv2-only -- Vector-only & -15.2 [-20.8, -10.2] & 1/49 & \(9.06\times10^{-14}\) & -0.199 [-0.250, -0.152] \\
\bottomrule
\end{tabular}%
}
\end{table}

Table~\ref{tab:retrieval-statistics} shows that Recall@20 differences involving AST-KG are not statistically significant and their target-cluster bootstrap intervals include zero. By contrast, the MRR@20 intervals for each reranked method minus vector-only exclude zero on the negative side. Under this analysis, vector-only has significantly better exact-target MRR than all three reranked methods.

Table~\ref{tab:retrieval-statistics} reports four external contrasts. The pre-specified controlled comparison is HippoRAG with document ensembling disabled versus ColBERTv2-only. The post-hoc comparisons quantify the effect of enabling document ensembling and the remaining difference between ensemble-enabled HippoRAG and ColBERTv2-only. The ColBERTv2-only versus Vector-only comparison characterizes the retrieval-backbone gap and is not used to infer the effect of graph augmentation. Paired comparisons of HippoRAG against vector-only and against both AST-KG conditions follow directly from Table~\ref{tab:retrieval-results} and are retained in the experimental artifacts rather than reproduced here, since they do not isolate the graph pipeline.

In the pre-specified HippoRAG comparison, discordance is strongly asymmetric, with 5 HippoRAG-only successes versus 161 ColBERTv2-only successes, so the deficit reflects a net loss rather than a redistribution of successes.

\subsection{Downstream Explanation Quality}

Table~\ref{tab:downstream-results} reports results on the 100-query downstream evaluation.

\begin{table}[htbp]
\caption{Downstream evaluation on 100 queries and 100 unique targets using \texttt{gpt-4.1-mini-2025-04-14}, top-five final-ranked evidence records, and single-evaluator blinded manual scoring. Lower hallucination is better; positive hallucination differences are worse. Discordant counts are \(A\)-only/\(B\)-only. Exact McNemar tests are the primary inference for binary paired outcomes; bootstrap intervals are complementary and need not invert the exact discrete test. Static-versus-staged comparisons are retained in the experimental artifacts.}
\label{tab:downstream-results}
\centering
\scriptsize
\resizebox{\ifdim\width>\linewidth\linewidth\else\width\fi}{!}{%
\begin{tabular}{llrrr}
\toprule
Metric & Condition / comparison & Value & Discordant & Exact \(p\) \\
\midrule
\multicolumn{5}{l}{\emph{Rates, \% [95\% Wilson interval]}} \\
RCA correctness & Vector-only & 58.0 [48.2, 67.2] & --- & --- \\
RCA correctness & AST-KG static & 57.0 [47.2, 66.3] & --- & --- \\
RCA correctness & AST-KG staged & 52.0 [42.3, 61.5] & --- & --- \\
Evidence grounding & Vector-only & 65.0 [55.3, 73.6] & --- & --- \\
Evidence grounding & AST-KG static & 69.0 [59.4, 77.2] & --- & --- \\
Evidence grounding & AST-KG staged & 63.0 [53.2, 71.8] & --- & --- \\
Hallucination & Vector-only & 54.0 [44.3, 63.4] & --- & --- \\
Hallucination & AST-KG static & 57.0 [47.2, 66.3] & --- & --- \\
Hallucination & AST-KG staged & 54.0 [44.3, 63.4] & --- & --- \\
\midrule
\multicolumn{5}{l}{\emph{Paired differences against Vector-only, pp [95\% paired-bootstrap CI]}} \\
RCA correctness & Static -- Vector & -1.0 [-7.0, +5.0] & 4/5 & 1.000 \\
RCA correctness & Staged -- Vector & -6.0 [-13.0, +1.0] & 4/10 & 0.180 \\
Evidence grounding & Static -- Vector & +4.0 [-4.0, +12.0] & 10/6 & 0.454 \\
Evidence grounding & Staged -- Vector & -2.0 [-11.0, +7.0] & 9/11 & 0.824 \\
Hallucination & Static -- Vector & +3.0 [-7.0, +13.0] & 15/12 & 0.701 \\
Hallucination & Staged -- Vector & 0.0 [-9.0, +9.0] & 11/11 & 1.000 \\
\bottomrule
\end{tabular}%
}
\end{table}
The 95\% Wilson intervals in Table~\ref{tab:downstream-results} are broad and overlap substantially across all three conditions and all three metrics.

The external HippoRAG experiment is restricted to retrieval; downstream generation was not rerun for it, so no external condition appears in Table~\ref{tab:downstream-results}.

Relative to vector-only, static AST-KG differs by \(-1\) percentage point in RCA correctness (\(p=1.000\)), while staged AST-KG differs by \(-6\) points (\(p=0.180\)). Grounding and hallucination comparisons likewise show no statistically reliable AST-KG advantage. Neither static nor staged AST-KG shows a statistically reliable improvement over vector-only in root-cause correctness, grounding, or hallucination, so the downstream study does not support improved diagnostic quality from AST-derived structural evidence. No equivalence, preservation, or non-inferiority claim is made.

\section{Discussion}
\label{sec:discussion}

Dense retrieval is a very strong baseline in this controlled setting. It obtains the highest MRR@\(K\) at every tested depth, and the target-cluster MRR@20 intervals show significantly worse exact-target rank for every reranked method. For Vector + CE, Recall@\(K\) is invariant by construction because the same dense candidate set is reranked; the reranker changed rank ordering and therefore MRR, improving 48 query ranks and worsening 79 relative to vector-only. Static AST-KG improved 38 and worsened 61 under the frozen rank-change definition. These observations caution against assuming that a reranker trained for general relevance will improve exact source-location ranking.

Local AST-derived method/class graph expansion also produces both rescues and losses. Relative to vector-only at \(K=20\), static AST-KG rescues four queries but loses seven. Three rescues concern the same \path{ReplicaManager.scala:2508} target, while five losses concern the same \path{FetchSessionHandler.java:618} target. Staged AST-KG has one rescue and one loss. This clustering illustrates why query-level counts alone can overstate patterns and why target-balanced metrics and target-cluster intervals are necessary. In the recorded static losses, the gold record is present in the expanded candidate pool but reranking moves it below rank 20; this bounds the explanation to the frozen artifacts without claiming a general causal mechanism.

The downstream evidence likewise does not show a statistically reliable RCA benefit. Static AST-KG is one point below vector-only correctness, staged AST-KG is six points below, the four-point static grounding increase is not significant, and neither AST-KG method reduces hallucination. Structural context may still be useful in other forms, but local ownership neighbors alone should not be assumed beneficial for incident diagnosis.

The external baseline shows that this absence of graph benefit is not peculiar to the AST-derived construction. Holding the retrieval backbone fixed at ColBERTv2, HippoRAG under its documented configuration achieves Recall@20 of 0.308 against 0.803 for direct retrieval, a paired difference of \(-49.5\) pp whose target-cluster interval excludes zero and whose discordance is 5 rescues against 161 losses.

The HippoRAG transfer result is configuration-sensitive in magnitude. Enabling document ensembling recovered \(+34.0\) pp of Recall@20 relative to the primary ensemble-off configuration and was associated with reduced ranking concentration. However, ensemble-enabled HippoRAG remained 15.6 percentage points below its matched ColBERTv2-only control. The evaluated external text-derived graph pipeline therefore did not improve exact source-location retrieval over its own backbone under either tested ranking mode. Neither the evaluated AST-derived graph conditions nor either HippoRAG condition establishes an exact-target retrieval advantage over the relevant non-graph baselines in this task. We did not demonstrate that graph-augmented retrieval is generally unhelpful; we observed that two specific graph constructions, one from program structure and one induced from record text, failed to beat their respective non-graph controls on this objective.

This result should not be interpreted as evidence that HippoRAG or text-derived graph retrieval is generally ineffective. HippoRAG was designed for knowledge integration and multi-hop question answering over prose-like passages, where the value of the graph lies in composing evidence spread across documents. Our task has the opposite shape: each retrieval unit is a short operational record, each query has exactly one correct \texttt{(file, line)} target, and success requires discriminating among 1,600 structurally similar records rather than composing several. We evaluated the configuration the documentation prescribes, without task-specific tuning, and varied exactly one documented option post hoc; the result therefore speaks to transfer under this configuration family and not to the method in its intended setting. That a single option recovered a third of the deficit is itself a caution against reading any one configuration's magnitude as characteristic of the system.

The diagnostics constrain which explanations remain available. The result is not readily explained by widespread extraction sparsity or graph fragmentation, nor by the uniform-ranking fallback, which affected only 4 of 315 queries, nor by query-side entity sparsity, since Hits@20 is approximately flat in the number of query entities. What the artifacts do show is a concentration effect together with a low-specificity vocabulary. HippoRAG with ensembling disabled returns a narrower slice of the corpus than ColBERTv2 alone, and the ensemble-enabled condition moves that coverage back toward the direct-retrieval condition. Meanwhile the most frequent extracted phrases are record-schema metadata and generic logging vocabulary rather than Kafka-specific entities. The frequent occurrence of serialization-related and generic logging phrases, together with more concentrated document rankings, is consistent with low-specificity graph connections distributing ranking mass across records. Our experiments do not isolate this mechanism as the sole cause, and no ablation was run that would license a causal claim.

The clear positive result is computational. Staged AST-KG equals or exceeds static AST-KG Recall at all tested depths and, at the primary \(K=20\), reduces cross-encoder pair workload by 51.8\% and measured cross-encoder reranking time by 51.4\%. These savings are relative to static AST-KG within one implementation. They are not end-to-end latency savings and do not establish a retrieval-quality advantage over vector-only.

This raises a fair question: if AST-KG does not improve retrieval quality, why does reducing its cost matter? The confidence rule itself depends on the cross-encoder score distribution rather than on AST-specific graph features. Although evaluated here only with AST-derived candidate expansion, the same gating principle could in principle be applied to other pipelines that expand a candidate pool before second-stage reranking, such as query expansion, hybrid retrieval, or graph-neighbor expansion. Whether comparable computational savings transfer to those settings remains future work; the present experiments establish the saving only for staged versus static AST-derived expansion. The negative quality result and the bounded efficiency result are both useful for AIOps/RAG design: structural expansion should be justified empirically, while confidence staging can reduce its cross-encoder pair workload and measured cross-encoder reranking time relative to static AST-KG. Neither the internal nor the external graph condition establishes a quality advantage over dense retrieval on this task, so the practical recommendation is to treat graph augmentation for exact source-location retrieval as a hypothesis to be tested per domain rather than as a default design choice.

\section{Threats to Validity}
\label{sec:threats}

The evaluation covers one Apache Kafka codebase and only 103 unique retrieval targets. Multiple queries refer to the same target, so nominal query count does not equal independent source-location count; target-balanced and target-cluster analyses mitigate but do not remove this limitation. Most queries are controlled incident-style descriptions derived from indexed source-linked records: the \texttt{v2} and \texttt{v3} subsets contribute 294 of the 315 queries, while 21 queries are JIRA-seeded. Neither group consists of complete production-triage transcripts. The JIRA-seeded queries cover only five unique source targets. Appendix~\ref{app:jira-robustness} reports an exploratory analysis restricted to this subset; its qualitative ordering is consistent with the main benchmark, but the strong target concentration prevents it from serving as an independent validation set or establishing that query provenance has no effect.

The downstream study uses one generator snapshot, \path{gpt-4.1-mini-2025-04-14}, and 100 unique targets. The 300 generated answers were scored manually by a single evaluator under blinded Method A/B/C identities using a fixed binary rubric. No independent second evaluator or inter-annotator agreement was available. The downstream design also omits a Vector+Cross-Encoder-only condition. Therefore, the results should be interpreted as a focused diagnostic evaluation rather than a multi-annotator benchmark.

The graph contains only local AST ownership among classes, methods, and supported logger calls in selected Java and Scala source areas. It does not encode control flow, data flow, call graphs, runtime traces, or operational topology. The parser recognizes configured logger tokens, levels, and literal forms; calls outside that scope may be omitted. Corpus counts and zero benchmark-target misses do not substitute for a labeled parser-accuracy gold set.

We evaluate one external text-derived graph-retrieval system, HippoRAG v1.0.0, but do not evaluate GraphRAG, GRAG, or other graph-RAG frameworks. The results reflect one Kafka corpus, 103 unique targets, and an exact source-location objective that differs from the multi-hop question-answering setting HippoRAG was designed for; they should not be generalized to that setting or to graph retrieval broadly. No domain-specific tuning of HippoRAG or ColBERTv2 was performed, so the comparison measures transfer of a prescribed configuration rather than the best achievable performance of either system. HippoRAG 2 introduces materially different passage-integration and reasoning mechanisms and was not evaluated.

The document-ensemble analysis was post-hoc, run after the primary result was observed, and only one option was varied. It demonstrates that the magnitude of the deficit is configuration-dependent; it does not establish that the ensemble-on setting is optimal, nor that other unexplored settings would not change the picture further. The primary ensemble-off condition remains the pre-specified comparison. Open information extraction was fixed at \texttt{gpt-3.5-turbo-1106}; a different extraction model could yield a different graph. Neither external condition was carried through to downstream generation.

The graph diagnostics establish that extraction was neither sparse nor fragmented, but connectivity does not establish the semantic specificity, quality, or task relevance of the extracted relations, and the reported concentration statistics are associations rather than a causal ablation. Our serialization audit further indicates that the most frequent extracted phrases are dominated by record-schema metadata and generic logging vocabulary, so the document serialization may itself influence which phrases open information extraction recovers; we did not vary the serialization to test this. Execution required platform and data-loading adaptations, including CPU-only execution on macOS, \texttt{faiss-cpu}, Python 3.10, sequential extraction, and minimum compatible LangChain dependency updates; these were not intended to alter retrieval or extraction semantics, but minor platform-dependent numerical differences remain possible. One further implementation detail affects the ensemble-on condition specifically: upstream's constructor rebuilds its document index through the stock loading path rather than through the compatibility adapter that preserves our serialization exactly. The previously measured difference between those two paths for ColBERTv2-only was 252/315 against 253/315 Hits@20. We report this as a minor implementation sensitivity rather than as proof of exact equivalence.

The equal weighting of the four confidence components and the 0.60 and 0.25 thresholds are heuristic design choices whose optimality this study does not establish.

Evaluation of HippoRAG 2 and additional graph-retrieval frameworks remains future work. Finally, cross-encoder pair counts and reranking time are measured in one implementation and environment. They exclude other retrieval, graph, network, and generation costs. The reported savings are reductions in cross-encoder pair workload and measured cross-encoder reranking time relative to static AST-KG, not universal or end-to-end latency savings. Downstream hallucination rates are likewise rubric- and sample-specific unsupported-causal-claim rates rather than universal model behavior.

\section{Conclusion}
\label{sec:conclusion}

This controlled evaluation finds no statistically reliable retrieval or RCA advantage for local AST-derived method/class graph expansion over vector-only retrieval. On 315 queries and 103 unique targets, vector-only remains strongest in exact-target MRR, and neither static nor staged AST-KG significantly improves Recall@20. The 100-query manually evaluated downstream study likewise finds no significant AST-KG root-cause-correctness advantage.

A controlled external experiment shows that this outcome is not specific to AST-derived graph construction. Evaluating HippoRAG on the identical frozen corpus and benchmark, and controlling for its ColBERTv2 retrieval backbone, HippoRAG under its documented configuration achieves Recall@20 of 0.308 compared with 0.803 for its matched ColBERTv2-only control. A post-hoc sensitivity analysis enabling HippoRAG's document-ensemble option materially improves transfer performance, raising Recall@20 to 0.648, but does not close the gap: ensemble-enabled HippoRAG remains 15.6 percentage points below its matched ColBERTv2-only control. The magnitude of the deficit is therefore configuration-sensitive, while its direction persists under the evaluated sensitivity condition, and neither graph-ranking mode improves over the backbone. Pre-specified diagnostics show that the extracted graph was neither sparse nor fragmented, so the result is not readily explained by extraction sparsity or fragmentation, although connectivity does not establish semantic specificity. This finding is task-specific. It concerns one configuration family transferred to short source-linked operational records with a single exact \texttt{(file, line)} target per query, and it is not a general indictment of graph-augmented retrieval or of HippoRAG in the multi-hop question-answering setting for which it was designed.

Relative to static AST-KG, confidence-guided staging nevertheless reduces cross-encoder pair workload and measured cross-encoder reranking time. At the primary \(K=20\), it retains the observed 301/315 hits while reducing cross-encoder pairs by 51.8\% and measured cross-encoder reranking time by 51.4\%. The \(K\)-sensitivity results support the same bounded efficiency interpretation. These findings caution against assuming that structural graph expansion automatically improves incident diagnosis. Future work should evaluate richer control-, data-, call-, runtime-, and topology-aware graphs; HippoRAG 2 and additional graph-retrieval frameworks; graph construction designed explicitly for operational and source-code records; additional software systems; and independent multi-annotator downstream evaluation.

\appendix
\makeatletter
\renewcommand\@seccntformat[1]{\appendixname~\csname the#1\endcsname\quad}
\makeatother
\section{JIRA-Seeded Query Robustness Analysis}
\label{app:jira-robustness}

Because 294 of the 315 benchmark queries come from the controlled \texttt{v2} and \texttt{v3} construction, in which prompts are generated from the indexed source-linked records themselves, we separately inspect the 21 JIRA-seeded queries as an exploratory robustness check on query provenance. These queries are phrased from public issue metadata rather than from the indexed records.

The subset is strongly target-concentrated and must be read with that in mind. The 21 queries represent only five unique expected source targets, with per-target query counts of 7, 6, 4, 3, and 1. It is therefore not an independent 21-case validation set, and query-weighted percentages over it are dominated by the two largest target groups. Table~\ref{tab:jira-subgroup} reports descriptive results; we deliberately omit inferential tests on this subset, because five effective clusters with heavy within-target repetition cannot support them.

\begin{table}[htbp]
\caption{Exploratory descriptive results on the 21 JIRA-seeded queries at \(K=20\). The subset covers only five unique expected source targets, so query-weighted values are dominated by the largest target groups; target-balanced Recall@20 gives each unique target equal weight. The ensemble-on HippoRAG row is a post-hoc sensitivity condition, not a primary condition. No inferential test is reported for this subset.}
\label{tab:jira-subgroup}
\centering
\resizebox{\ifdim\width>\linewidth\linewidth\else\width\fi}{!}{%
\begin{tabular}{lrrrr}
\toprule
Method & Hits@20 & Recall@20 & tb-Recall@20 & MRR@20 \\
\midrule
Vector-only & 21/21 & 1.000 & 1.000 & 0.630 \\
Vector + CE & 21/21 & 1.000 & 1.000 & 0.637 \\
AST-KG static & 16/21 & 0.762 & 0.857 & 0.629 \\
AST-KG staged & 21/21 & 1.000 & 1.000 & 0.637 \\
ColBERTv2-only & 16/21 & 0.762 & 0.857 & 0.605 \\
HippoRAG (ensemble off) & 5/21 & 0.238 & 0.329 & 0.016 \\
HippoRAG (ensemble on, post-hoc) & 8/21 & 0.381 & 0.490 & 0.048 \\
\bottomrule
\end{tabular}%
}
\end{table}

Vector-only retrieves all 21 JIRA-seeded targets within \(K=20\), and staged AST-KG also retrieves all 21; staged AST-KG therefore shows no JIRA-only Recall@20 advantage over vector-only retrieval. Static AST-KG retrieves 16 of 21. Crucially, all five static-versus-vector discordances correspond to the same expected target, \path{FetchSessionHandler.java:618}, which alone accounts for seven of the 21 queries. The apparent five-query difference therefore reflects one clustered source-location case rather than five independent target failures, and the same target accounts for all five ColBERTv2-only misses. Target-balanced Recall@20 is 1.000 for vector-only and staged AST-KG and 0.857 for static AST-KG. One isolated counter-case runs the other way: query \texttt{KAFKA-9357-Q3} is ranked 6 by static AST-KG and 17 by staged AST-KG.

For the external conditions, ColBERTv2-only retrieves 16 of 21, HippoRAG with document ensembling disabled retrieves 5, and the post-hoc ensemble-enabled condition retrieves 8. The qualitative external ordering, ColBERTv2-only above ensemble-enabled HippoRAG above ensemble-disabled HippoRAG, is consistent with the full benchmark.

Overall, the issue-seeded subset is qualitatively consistent with the main finding that AST-derived structural expansion does not provide a clear exact-target retrieval advantage over dense retrieval. However, because the 21 queries represent only five unique targets, this analysis is an exploratory query-provenance robustness check rather than an independent validation of the primary result.

\section*{Acknowledgment}

The authors thank the Apache Kafka open-source community for the codebase and operational logging patterns that made this evaluation possible.

\section*{Author Biographies}

\noindent\textbf{Shashank Mishra} received the M.S. degree in computer science from the University of Southern California, Los Angeles, CA, USA, in 2020. He is currently an independent researcher based in Seattle, WA, USA. His research interests include large-scale distributed training, memory-efficient systems for large language models, retrieval-augmented generation, knowledge-graph-augmented retrieval, LLM evaluation, and applied AIOps for incident diagnosis in distributed systems.

\medskip

\noindent\textbf{Shweta Mishra} received her graduate degree from Boston University, Boston, MA, USA, and is currently an independent researcher based in San Jose, CA, USA. Her research interests include artificial intelligence systems, large language models, and intelligent agent architectures, with a focus on memory architectures for AI agents, retrieval-augmented generation, multi-agent frameworks, and distributed LLM infrastructure for scalable, efficient foundation-model applications.

\end{document}